\documentclass[dvipdfmx]{PoS}
\usepackage{wrapfig}
\usepackage{amsmath,amssymb,amsthm,bm,slashed,ascmac,emathCap}
\usepackage{multicol}
\newcommand{\PM}{P_{-}}
\newcommand{\PP}{P_{+}}
\newcommand{\PR}{P_{R}}
\newcommand{\PL}{P_{L}}

\title{The one-loop analysis of the beta-function in the Schroedinger Functional for Moebius Domain Wall Fermions}

\ShortTitle{The one-loop analysis of the beta-function in the SF for MDWF}

\author{\speaker{Yuko Murakami}\\%\thanks{A footnote may follow.}\\
        Graduate school of Science, Hiroshima University, Higashi-Hiroshima, Hiroshima 739-8526, Japan\\
        E-mail: \email{yukomhir@gmail.com}}

\author{Ken-Ichi Ishikawa\\
        Graduate school of Science, Hiroshima University, Higashi-Hiroshima, Hiroshima 739-8526, Japan,\\
        Core of Research for Energetic Universe, Hiroshima University, Higashi-Hiroshima, Hiroshima 739-8526, Japan\\
        E-mail: \email{ishikawa@theo.phys.sci.hiroshima-u.ac.jp}}

\abstract{
We proposed a construction of the Schroedinger functional scheme 
for the Moebius domain wall fermions (MDWF) by introducing a proper boundary operator 
to the original MDWF in the last conference. 
The spectrum of the effective four-dimensional operator was investigated.
This year we investigate the fermionic contribution to the beta-function 
with the Moebius domain wall fermion with the SF boundary term up to the one-loop level
and find that our construction properly reproduce the one-loop beta-function.}

\FullConference{The 33rd International Symposium on Lattice Field Theory\\
		14 -18 July 2015\\
		Kobe International Conference Center, Kobe, Japan*}

\begin{document}
%%%%Section : Introduction
\section{Introduction}
The chiral symmetry is important not only in QCD but also in the standard model.
To extract hadronic observables based on the chiral symmetry, we need both the lattice field theory and 
the chiral symmetry. The lattice chiral symmetry is now realized by the overlap and domain wall fermions 
and extensive efforts have been devoted to the accurate values for hadronic observables sensitive 
to the chiral symmetry\cite{Flynn:2015mha,Aoki:2013ldr}. 
In this sense the non-preturbative renormalization for the lattice fermions with the lattice chiral 
symmetry becomes important.

The Schroedinger functional (SF) scheme~\cite{Luscher:1985} has been successfully applied to renormalize the lattice QCD
with the Wilson fermions non-perturbatively~\cite{Luscher:1993gh}.  
Although it was not simple for the lattice chiral fermions to apply the SF scheme, 
the SF methods for the lattice chiral fermions have been developed in Refs.~\cite{Luscher:2006df,Taniguchi:2004gf,Taniguchi:2006qw,Takeda:2010ai}.

In the last lattice conference, we proposed a construction of the SF scheme for Moebius domain 
wall fermions (MDWF)~\cite{Murakami:2014qfa}.
The operator is modified by adding a temporal boundary operator based on Takeda's implementation for the standard domain wall 
fermion~\cite{Takeda:2010ai}. 
The form of the boundary operator is determined by the symmetry argument of 
Ref.~\cite{Luscher:2006df} so that
the term holds the discrete symmetries ($C,P,T,\Gamma_5$-Hermiticity) and breaks
the domain wall chiral symmetry~\cite{Furman:1994ky} at the temporal boundary.
We investigated the properties of the effective four-dimensional operator induced from 
the MDWF with the boundary operator and found that it reproduced the proper continuum theory 
with the SF boundary only in the infinite fifth lattice extent $N_5\to \infty$.

In this paper, we extend our previous work to obtain the proper operator for the SF scheme even at a finite $N_5$.
To do this we properly renormalize the effective four-dimensional operator 
since the explicit breaking of the lattice chiral symmetry at a finite $N_5$ induces the additive 
residual mass~\cite{Capitani:2007ez} as seen in Wilson type fermions even at the tree-level. 
After renormalizing the effective four-dimensional operator we investigate the eigenvalues in the SF boundary condition.
In order to check the consistency of our construction, we also investigate the fermionic contribution to 
the one-loop beta-function in the SF scheme at a finite $N_5$.

This paper is organized as follows.
In the next section, we briefly explain the renormalization factors for the MDWF 
at the tree-level without the SF boundary condition. The renormalization factors are applied to the case of the SF scheme. 
Then we introduce the MDWF with the SF boundary term given by our previous work~\cite{Murakami:2014qfa}. 
The lowest eigenvalues of the effective four-dimensional operator are investigated.
In section~\ref{sec:oneloop}, we confirm that the properly renormalized MDWF operator with the SF boundary term
 reproduces the proper one-loop beta-function. 
The scaling violation on the step scaling function is also examined at the one-loop level. 
We summarize this work in the last section. 

%%%%Section : Set up
\section{The set up of the SF scheme for the MDWF}
\label{sec:set_up}
The Moebius domain wall fermion (MDWF)~\cite{Brower:2004xi} is one of the lattice chiral fermion
and defined as the fifth dimensional operator as follows.
\begin{align}
 D_{\mathrm{MDWF}}(n,s;m,t) &= 
 D_s^{+} (n;m) \delta_{s,t} + D_s^{-} (n;m) M_{5}(s;t),\\
   D_s^{+}(n;m) &= b_s D_{\mathrm{WF}} (n;m) + 1, \\
   D_s^{-}(n;m) &= c_s D_{\mathrm{WF}} (n;m) - 1, \quad (s=1,\cdots,N_5) \\
  M_{5}(s;t) &= \PL \delta_{s+1,t} + \PR \delta_{s-1,t}
  -m_f [ \PL \delta_{s,t} \delta_{1,t} + \PR \delta_{s,1} \delta_{N_5,t}],
\end{align}
where $(s,t)$ is the lattice index in the fifth direction, $(n,m)$ is the four-dimensional 
lattice site index,
$m_f$ is the mass parameter, $D_{WF}$ is the Wilson fermion operator with a negative mass ($m_0$),
$P_{R/L}$ is the chiral projection: $P_{R/L} = (1 \pm \gamma_5)/2$, 
$(b_s, c_s)$ are the Moebius parameters, and $N_5$ is the lattice size in the fifth direction. 
The MDWF is a generalization of the domain wall fermions and includes the standard domain wall fermion (SDWF)~\cite{Furman:1994ky},
the Borici domain wall fermion (BDWF)~\cite{Borici:1999da}, the optimal Shamir domain wall fermion~\cite{Borici:2002xe}
and the optimal Chiu domain wall fermion (CDWF)~\cite{Chiu:2002ir} by adjusting $(b_s,c_s)$.
The Moebius parameters for the optimal type domain wall fermions can be derived 
from the Zolotarev sign function approximation~\cite{Chiu:2002eh}.

The effective four-dimensional operator calculated from the MDWF operator yields the overlap fermion operator
~\cite{Narayanan:1994gw,Neuberger:1997fp}
 at the infinite extent of the fifth direction~\cite{Kikukawa:1999sy, Borici:2004pn} ($N_5\to\infty$). 
At a finite $N_5$ the effective operator does not satisfy the Ginsparg-Wilson relation~\cite{Ginsparg:1982}
and an $O(a)$ error is expected, so that
the relation to the pole mass and the bare mass $m_f$ can be differ in this case.
The effective four-dimensional operator, $D_{\mathrm{eff}}^{(N_5)}$, at a finite $N_5$ 
without the SF boundary condition is evaluated as
\begin{align}
  aD_{\mathrm{eff}}^{(N_5)} & 
\equiv
P^T D_{\mathrm{PV}}^{-1} D_{\mathrm{MDWF}}(am_f) P
\notag \\
    &= \frac{1 + am_f}{2} - \frac{1 - am_f}{2} \gamma_5 R_{N_5}(\mathcal{H}_{W}),
\label{eq:effopwoSF}
 \\
  R_{N_5}(\mathcal{H}_{\omega})
    &=
  \frac{  \prod_{s=1}^{N_5} (1 + \omega_{s}\mathcal{H}_{W})
        - \prod_{s=1}^{N_5} (1 - \omega_{s}\mathcal{H}_{W})}
       {  \prod_{s=1}^{N_5} (1 + \omega_{s}\mathcal{H}_{W})
        + \prod_{s=1}^{N_5} (1 - \omega_{s}\mathcal{H}_{W})}, \\
  \mathcal{H}_{W} &= \frac{aD_{WF}}{a_5D_{WF}+2}, 
\label{eq:KernelOP}
\\
  a_5 &= b_s-c_s,\quad\quad
  \omega_s = b_s+c_s,\\
P(s) &= \PL \delta_{s,1} + \PR \delta_{s,N_{5}},
\label{eq:effov}
\end{align}
where $D_{\mathrm{PV}} = D_{\mathrm{MDWF}}(am_f=1.0)$.
The coefficients $a_5$ and $\omega_s$ 
must be tuned to properly reproduce the sign function as
$R_{N_5}(x) \to \mathrm{sign}(x)$ in the limit of $N_5\to\infty$. 
The ordering of $(b_s,c_s)$ in the fifth direction is irrelevant for 
the effective four-dimensional operator without the SF boundary condition.
In the continuum limit 
the effective four-dimensional operator behaves as
\begin{align}
  a D_{eff}^{(N_5)} &\approx Z_{N_5} (a \slashed{D} + a m_{\mathrm{res}}), 
  \label{eq:releffov}\\
  Z_{N_5} &= \dfrac{(1-am_f) R_{N_5}(\alpha)}{(am_0)(2-(am_0)a_5)}, \\
 am_{\mathrm{res}} &= \left[\dfrac{1+am_f}{1-am_f}\dfrac{1}{R_{N_5}(\alpha)} -1\right]\dfrac{(am_0)(2-(am_0)a_5)}{2}, 
\label{eq:residualmass}
\\
  \alpha & = \dfrac{(am_0)}{(2-(am_0)a_5)}, 
\end{align}
where $\slashed{D}$ is the Dirac operator in the continuum, 
$Z_{N_5}$ is the normalization factor and 
$m_{\mathrm{res}}$ is the residual mass at the tree-level~\cite{Capitani:2007ez}. 
Even at $m_f=0$ the residual mass does not vanish 
as $R_{N_5}(\alpha) \ne 1$ at a finite $N_5 < \infty$.
In order to investigate the property of the effective operator at the vanishing pole mass,
we renormalize the effective operator as follows,
\begin{align}
  a D_R^{(N_5)} &= Z_{N_5}^{-1} aD_{eff}^{(N_5)}(am_{\mathrm{cr}}),  \quad \mbox{where} \quad
  am_{\mathrm{cr}} = \frac{R_{N_5}(\alpha)-1}{R_{N_5}(\alpha)+1}.
  \label{eq:def_reno_op}
\end{align}

Now we consider the effective four-dimensional operator in the SF scheme.
In our previous work, we proposed the following operator for the MDWF in 
the SF scheme~\cite{Murakami:2014qfa}.
\begin{align}
D_{\mathrm{MDWF}}^{\mathrm{SF}} (n,s;m,t) &= 
D_{\mathrm{MDWF}}(n,s;m,t) + c_{\mathrm{SF}}B_{\mathrm{SF}}(n,s;m,t)
\label{eq:MDWFSF},
\\
B_{SF}(n,s;m,t)&= 
   f(s)
   \delta_{s,N_5-t+1}
   \delta_{\bm{n},\bm{m}} \delta_{n_4,m_4}
   \gamma_5 (\delta_{n_4,1} \PM + \delta_{n_4,T-1} \PP) ,
\\
f(s) & =\left\{
\begin{matrix}
 -1 & (\mbox{for $1\le s \le N_5/2$}) \\
 +1 & (\mbox{for $N_5/2+1 \le s \le N_5$})
\end{matrix}\right.,
\end{align}
where
$P_{\pm}$ is the projection: $P_{\pm}=(1\pm\gamma_4)/2$.
We introduce the boundary operator $B_{\mathrm{SF}}$ to satisfy 
the SF boundary condition, which explicitly breaks the domain wall chiral 
symmetry~\cite{Furman:1994ky} at the temporal boundary, 
as suggested in Refs.~\cite{Luscher:2006df,Takeda:2010ai}.
In order to keep the discrete symmetries $C,P,T$,and $\Gamma_5$-Hermiticity
we require the parity symmetry in the fifth direction because 
the ordering of the coefficients $(b_s,c_s)$ in the fifth direction 
is relevant in this case.
A quasi optimal choice for $\omega_s$ with the parity symmetry are determined 
according to Ref.~\cite{Murakami:2014qfa}.
We call the domain wall fermion operators with the quasi optimal 
choice for $\omega_s$ having the parity symmetry in the fifth direction 
as 
the palindromic Shamir domain wall fermion (PSDWF) 
and 
the palindromic Chiu domain wall fermion (PCDWF)
depending on the choice for the kernel operator.
We use the same form as Eq.~(\ref{eq:def_reno_op}) for the normalization factor 
and the pole mass with the SF boundary condition, though we cannot obtain the effective 
four-dimensional operator in a simple closed form like Eq.~(\ref{eq:effopwoSF}) 
and the corresponding behavior in the continuum limit like Eq.~(\ref{eq:releffov}).

\begin{wrapfigure}{r}{0.47\linewidth}
\vspace*{-2em}
    \begin{center}
  \includegraphics[clip,scale=0.34,trim=10 205 0 210]{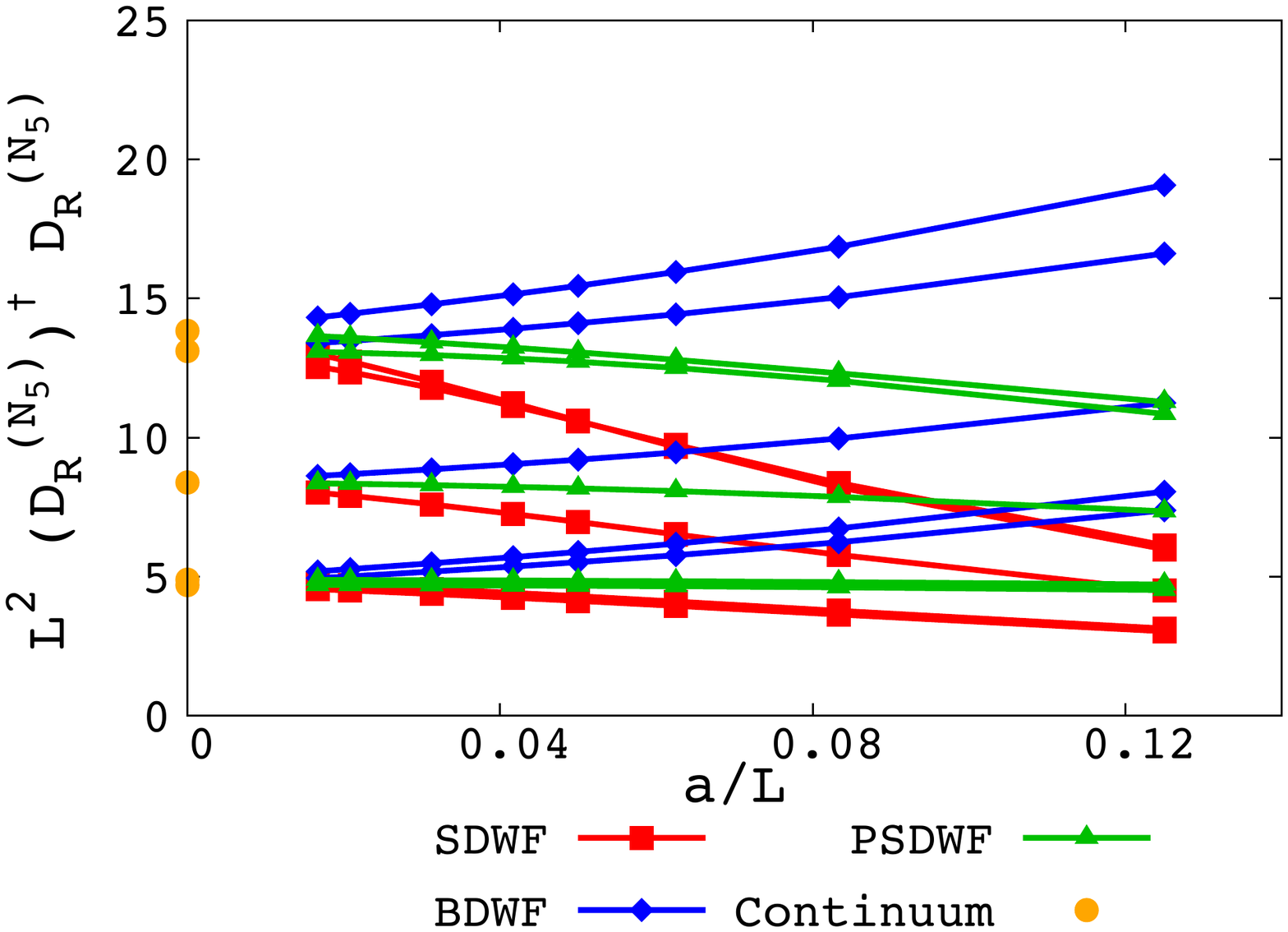}
\vspace*{-1em}
  \caption{The scaling behavior of the lowest five eigenvalues of 
             the Hermitian squared effective four-dimensional operator.}
  \label{fig:evtree_r}
\vspace*{-2.5em}
    \end{center}
\end{wrapfigure}

We investigate the lowest eigenvalues of the effective four-dimensional operator 
derived from Eq.~(\ref{eq:MDWFSF}) at the tree-level.
We use $L=T$ and the standard boundary condition for the gauge field~\cite{Luscher:1993gh}
yielding a constant background chromoelectric field.
The generalized periodic boundary condition with the phase angle $\theta=\pi/5$ is used for the spatial boundary condition.
In order to study the effect of the renormalization
with a finite fifth dimensional extent at the tree-level,
we set $m_0=1.5$ and $(b_s,c_s) = (1.0,0.0)$ for the SDWF 
   and $m_0=1.5$ and $(b_s,c_s) = (1.0,1.0)$ for the BDWF 
   and $m_0=1.0$ for the PSDWF at $N_5=8$ and $c_{SF} = 1.0$.

%\newpage

Figure~\ref{fig:evtree_r} shows 
the lowest five eigenvalues of the Hermitian squared effective four-dimensional operator 
$L^2 (D_R^{(N_5)})^{\dag}D_R^{(N_5)}$ for each operator type.
The yellow dot points at the continuum limit are quoted from Ref~\cite{Sint:1995ch}.
We confirm that the operator
involving the tree-level renormalization  properly reproduces the continuum theory.

%%%%Section : The one-loop beta-function
\section{The one-loop beta-function}
\label{sec:oneloop}
We calculate the fermoinic contribution to the one-loop beta-function using 
the MDWF with the SF boundary condition.
The renormalized coupling constant in the SF scheme~\cite{Luscher:1985} is defined by
\begin{align}
  \frac{1}{g^2_{\mathrm{SF}}} &= 
\left.\dfrac{1}{k}\frac{\partial \Gamma}{\partial \eta} \right|_{\eta=\nu=0},
\end{align}
where $\Gamma$ is the effective action with the SF boundary condition, 
$\eta$ and $\nu$ are parameters for the SF boundary condition, 
and $k$ is a normalization factor based on the tree-level analysis.
For the MDWF in the SF scheme, the fermionic contribution to the one-loop 
beta-function can be obtained from the following term;
\begin{align}
  p_{1,1} & = \left.\frac{1}{k} \dfrac{\partial \Gamma_1}{\partial \eta} \right|_{\eta=\nu=0}
    = 
\left.
\dfrac{1}{k} \frac{\partial }{\partial \eta} 
\left[ \ln{\mathrm{det} (D_{\mathrm{MDWF}}^{\mathrm{SF}} (D_{\mathrm{PV}}^{\mathrm{SF}})^{-1}) }\right]\right|_{\eta=\nu=0},
\end{align}
where $\Gamma_1$ is the one-loop effective action of the fermion part and 
$D_{\mathrm{PV}}^{\mathrm{SF}}=D_{\mathrm{MDWF}}^{\mathrm{SF}}(am_f=1.0)$.
The cut-off dependence of $p_{1,1}$ can be written in the following form asymptotically;
\begin{align}
  p_{1,1} 
    &= \sum_{n=0}(a/L)^n [A_n + B_n \ln{(L/a)}] 
      = A_0 + B_0 \ln{(L/a)} + (a/L)(A_1 + B_1 \ln{(L/a)}) +O(a^2).
  \label{eq:p11}
\end{align}
$B_0$ contains the information of the fermionic contribution to the one-loop beta-function 
and it should be $B_0=2b_{0,1}=-1/(12 \pi^2)=-0.008443 \ldots$.
We numerically calculate $p_{1,1}$ using the MDWF with the SF boundary of Eq.~(\ref{eq:MDWFSF}) 
and the proper renormalization at the vanishing pole mass,
and fit them with the function Eq.~(\ref{eq:p11}) up to $O(a)$.
Table~\ref{tab:beta} shows the result of the fitting. 
The values for $c_{\mathrm{SF}}$ are determined with
the $O(a)$ improvement procedure based on the PCAC relation~\cite{Luscher:1996sc} 
at the tree-level.
We find that $B_0$ are consistent with the $2 b_{0,1}=-0.008443 \ldots$ 
within the fitting error for each DWF type and confirm that
the MDWF with the SF boundary term we constructed at the tree-level
properly reproduces with the fermionic contribution to the one-loop beta-function. 

\begin{table}[htb]
  \centering
  \begin{tabular}{|c|c|c|c|c|c|}\hline
  Operator & $m_0$ & $c_{SF}$ & $N_5$ & Fit range $[L_{min}/a,L_{max}/a]$& $B_0 \quad (\times 10^{-3})$   \\\hline
  SDWF   & $1.5$ & $0.520$ & $ 8$  & $[18:48]$ & $-8.43  \pm 0.01 $ \\\cline{4-6}
               &            &               & $16$ & $[26:48]$ & $-8.441 \pm 0.004 $ \\\hline  
  BDWF   & $1.5$ & $0.312$ & $ 8$  & $[18:48]$ & $-8.37  \pm 0.08  $ \\\cline{4-6}
               &           &               & $16$ & $[34:48]$ & $-8.46  \pm 0.05  $ \\\hline
 PSDWF & $1.0$ & $0.820$ & $ 8$  & $[10:48]$ & $-8.2   \pm 0.3   $ \\\cline{3-6}
               &           & $0.630$ & $16$ & $[16:48]$ & $-8.44  \pm 0.03  $ \\\hline  
  \end{tabular}
  \caption{The fit results for $B_0$.}
  \label{tab:beta}
\end{table}

We investigate the lattice cut-off dependence for the step scaling function~(SSF)~\cite{Luscher:1991}.
The deviation of the SSF at a finite cut-off from that in the continuum limit 
is defined by
\begin{align}
  \delta &= \dfrac{\Sigma (s,u,a/L) - \sigma(s,u)}{\sigma(s,u)}
    = \delta_1 u + \delta_2 u^2 + \cdots,
\end{align}
where 
$u$ is the SF scheme coupling constant $u=g^2_{\mathrm{SF}}(L)$ renormalized at $L$,
$\Sigma (s,u,a/L) = g^2_{\mathrm{SF}}(sL)$ is the SSF at a finite lattice cut-off, 
    $\sigma (s,u) = g^2_{\mathrm{SF}}(sL)$ is that in the continuum theory, and $s$ is a scale factor.
$\delta$ can be expanded as a polynomial of $u$ perturbatively as shown in 
the equation. 
$\delta_{1}$ must vanish in the continuum limit at the one-loop level analysis.
The fermionic contribution to $\delta_{1}$ is given by
\begin{align}
  \delta_1 &= \delta_{1,0} + \delta_{1,1}n_f, \\
  \delta_{1,1} &= p_{1,1}(a/(2L)) - p_{1,1}(a/L) - 2b_{0,1} \ln{2},
  \label{eq:drb_ssf}
\end{align}
where we set $s=2$.

\begin{figure}[t]
  \centering
  \includegraphics[width=80.0mm,clip,trim=0 200 0 200]{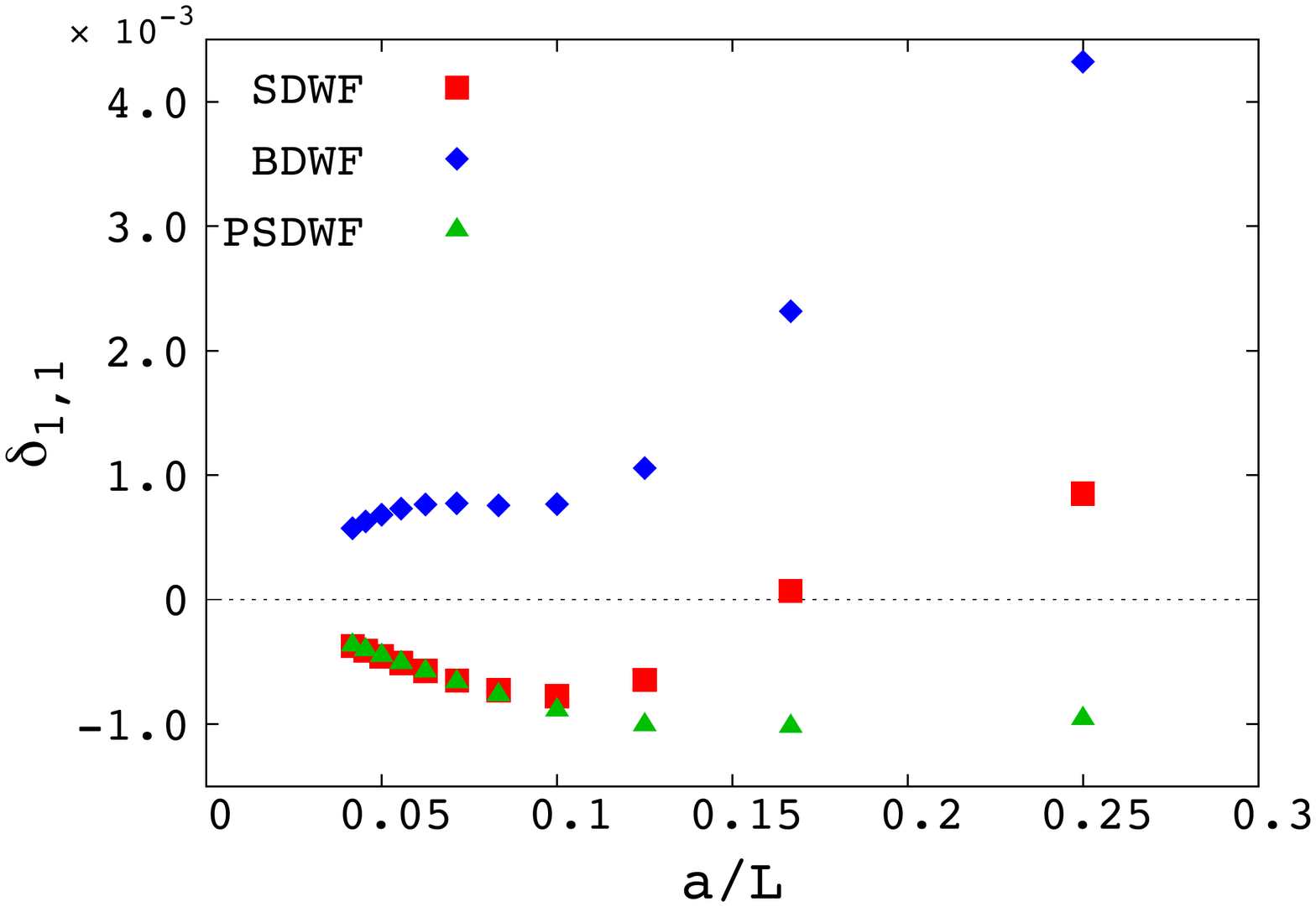}
  \vspace*{-1em}
  \caption{The deviation of the SSF between the lattice and the continuum 
           theory as a function of the lattice spacing. ($N_5=16,2 b_{0,1}=-1/(12\pi^2)$)}
  \label{fig:ssf}
\end{figure}

The lattice cut-off dependence of Eq.~(\ref{eq:drb_ssf})
with the MDWF at $N_5=16$ is shown in Figure~\ref{fig:ssf}, 
where $2b_{0,1}=-1/(12\pi^2)$ is imposed.
Although rather complicated cut-off dependence of $\delta_{1,1}$ is seen,
it goes to vanish in the continuum limit. 
Therefore we conclude that the renormalized MDWF with the SF scheme is consistent with 
the continuum massless Dirac operator at the one-loop level.
The cut-off dependence includes the $O(a)$ error from $A_1$ and $B_1$ terms 
of Eq.~(\ref{eq:p11}).
The former can be removed by tuning the boundary operator of the gauge action, while
the latter requires the bulk $O(a)$-improvement term such 
as the clover term for the Wilson fermion because
the lattice chiral symmetry is broken explicitly at a finite $N_5$.

%%%%
\section{Summary}
\label{sec:summary}
We investigated the Schroedinger Functional (SF) scheme with the Moebius domain wall fermions (MDWF). 
The MDWF with the SF boundary term was introduced and the spectrum of the massless Dirac operator 
with the SF boundary condition is reproduced from the MDWF at the tree-level 
after applying the proper renormalization even at a finite fifth dimensional extent.
We also confirmed that the MDWF operator with the SF boundary term reproduced 
the fermionic part of the universal one-loop beta-function in the SF scheme.
From these analysis at the one-loop level, we expect that the SF scheme is applicable to the 
MDWF at a finite $N_5$, which can be regarded as a kind of better Wilson fermions, 
by adding the SF boundary term.

We will check the consistency of the step scaling function of the coupling 
calculated from the MDWF with the SF boundary term non-perturbatively 
in the future work.

\section*{Acknowledgment}
We thank to S.~Takeda for the helpful advice. A part of numerical computations is performed on 
the INSAM (Institute for Numerical Simulations and Applied Mathematics) GPU Cluster at Hiroshima 
University.
This work was supported in part by a Giant-in-Aid for Scientific Research (C) (No. 24540276) 
from the Japan Society for the Promotion of Science (JSPS) and the MEXT program for promoting 
the enhancement of research universities, Japan.

\end{document}